\documentclass[a4paper,12pt]{article}

\usepackage{amsmath}
\usepackage[final]{graphicx}
\usepackage{bm}
\usepackage{amssymb}

\usepackage{amssymb}
\newcounter{comment}

{\refstepcounter{comment}%
\begin{quote}
\ttfamily\small$\blacksquare$ \textbf{\underline{Comment} $\sharp$\thecomment:}}%
{\end{quote}}

\begin{document}
\hfill
\begin{minipage}{20ex}\small
OSLO-TP-2-05\\
ZAGREB-ZTF-05-01\\
\end{minipage}

\begin{center}
\baselineskip=2\baselineskip
\textbf{\LARGE{
Soft gluon contributions to 
the $\boldsymbol{B\to K\eta'}$ amplitude 
in a low energy bosonization model
}}\\[6ex]
\baselineskip=0.5\baselineskip

{\large Jan~O.~Eeg$^{a,}$\footnote{j.o.eeg@fys.uio.no; corresponding author},  
Kre\v{s}imir~Kumeri\v{c}ki$^{a,b,}$\footnote{kkumer@phy.hr}, and
Ivica~Picek$^{b,}$\footnote{picek@phy.hr}}\\[4ex]
\begin{flushleft}
\it
$^{a}$Department of Physics, University of Oslo, P.O.B. 1048 Blindern, N-0316 Oslo, 
Norway\\[1.5ex]
$^{b}$Department of Physics, Faculty of Science, University of Zagreb,
 P.O.B. 331, HR-10002 Zagreb, Croatia\\[3ex]
\end{flushleft}
\today \\[5ex]
\end{center}

\begin{abstract}

Intriguing $B\to K\eta'$ decays provide a unique opportunity to study a
joining of  two-gluon  configurations arising from the penguin
$b\to s G$ and $b\to s  GG$ transitions, with those inherent to the $\eta'$ particle. 
We employ the heavy-light chiral quark model, applied previously to a somewhat
related  $B\to D\eta'$ decay, as a calculational tool accounting for the
nonperturbative soft gluon contributions to the amplitude at hand. Thereby we
arrive at a novel contribution to the singlet penguin amplitude, which within
our model  accounts for $\sim 10\%$ of the measured $B\to K\eta'$ amplitude.

\end{abstract}

\vspace*{2 ex}

\begin{flushleft}
\end{flushleft}

\clearpage

\section{Introduction}

The data on rare decays from CLEO, BaBar, Belle and Tevatron hold promise for
deepening our understanding of the interplay between flavour-changing and QCD
dynamics. In particular, the surprise regarding the $B\to K\eta'$ amplitude,
existing already for considerable time \cite{bketaexper,HFAG},
calls for an explanation.
Namely, the branching ratio for the decay mode $B^{+}\to K^{+}\eta'$ is measured to be
almost six times bigger than the one
for $B^{+}\to K^{+} \pi^0$, although the same basic
 $b\to s$ penguin mechanism is expected to drive both processes.
Apparently, this mechanism may have a different appearance when
instead
 of the flavour octet pion there is an (almost) flavour singlet $\eta'$ particle involved.
 
It is well known that some extraordinary properties of the $\eta'$ particle
are related to the QCD anomaly. Therefore,  a suggestion to explain the enhancement
of the $B\to K\eta'$ amplitude by the QCD anomaly at first sight looks very
intriguing \cite{bketatheor}.

 Our preceding investigation in this direction found that 
the $b\to s\eta'$ amplitude in the hard gluon regime represents a well
defined short distance (SD)  mechanism \cite{EeKP03}, but of minor numerical importance
for the process at hand.
The inability of this SD mechanism to account for the measured amplitude
invites us to explore here the complementary long-distance (LD) mechanism.
There are conclusions
in the literature \cite{Chiang:2003rb} that the ``singlet
penguin'' amplitude (in the language of SU(3) diagrammatic approach \cite{diagramsu3})
 contributes substantially to the $B\to K\eta'$ enhancement. One of the
purposes of this note is to investigate this from another angle than done
previously, i. e. to identify the singlet penguin contribution and
estimate it within the well-defined
microscopic dynamical framework.

In order to deal with the low energy properties of some of the
involved gluons and to calculate the physical 
$B\to K\eta'$ amplitude, we have to introduce an appropriate 
low-energy description.
 We will employ ideas based on the chiral quark
model ($\chi$QM)\cite{ChQM,ERT} which has been used to describe 
$K \rightarrow 2 \pi$ decays \cite{PideR,BEF}.
Here we will rely on the extension of such models, namely the
 heavy-light chiral quark model (HL$\chi$QM) \cite{Hiorth:2002pp}
which has, as the $\chi$QM,  turned out to be
 a convenient calculational tool for addressing soft-gluon contributions, 
 that can be expressed in terms of gluon condensate effects
 \cite{ERT,PideR,BEF,Hiorth:2002pp,Nov}.
One should note that when the quarks and soft gluons in  the
 HL$\chi$QM are integrated out, one obtains standard heavy light
 chiral perturbation theory (HL$\chi$PT) \cite{HLChPT}.
The  HL$\chi$QM   has been applied to $B - \bar{B}$
mixing \cite{Hiorth:2003ci} and to decays of
 the type $B \rightarrow D \bar{D}$ \cite{EFH}. 
It has also been applied to the decay mode $B \rightarrow D \eta'$ \cite{Eeg:2001tg}
which has similar aspects as the mode we consider in this paper. 
In principle, the HL$\chi$QM naturally accounts only for soft kaons in the final state.
However, as we show in the next section, assuming the general form of
the relevant  form factor enables one to
perform extrapolation from the soft to the hard kaon case.

\section{Penguin contribution to $\boldsymbol{B\to K\eta'}$ 
in the heavy-light chiral quark model}

\begin{figure}
\centerline{\includegraphics[scale=1.4]{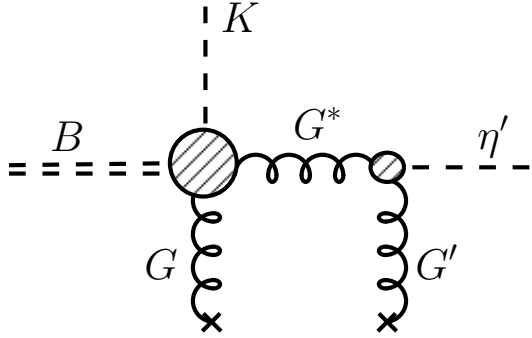}}
\caption{\label{BKetaGG}{A mechanism for $B\to K \eta'$ due to soft
    gluons. The crosses
correspond to the gluon condensate.}}
\end{figure}

\begin{figure}
\centerline{\includegraphics[scale=1.2]{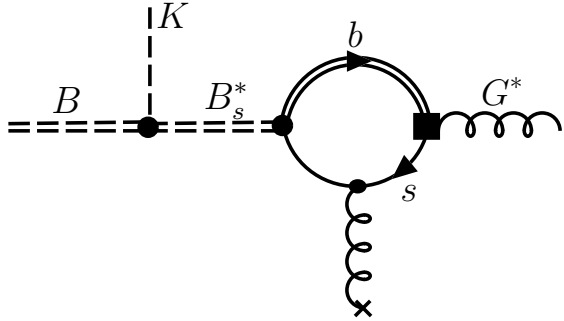}}
\caption{\label{BKGGst}{$B \rightarrow K G G^*$ in the soft kaon
    limit. The black box denotes the $b\to s G^*$ penguin transition.}}
\end{figure}

The LD mechanism that we propose for the $B\to K\eta'$ decay is shown
in Figure~\ref{BKetaGG}. It accounts for the contribution obtained when a 
soft gluon ($G$) is emitted from the $B\to K$ transition together with
the virtual gluon ($G^*$) associated with the penguin $b \rightarrow
s$ transition. In addition, one
of the gluons ($G'$) in the $\eta' GG$ vertex
in Figure~\ref{BKetaGG}
is also assumed to be soft,
so that two soft gluons form a vacuum condensate.
The remaining  off-shell gluon from the $b\to s G^*$ penguin is
propagating into the $\eta'$. 
Now, the $B\to K G G^*$ vertex denoted by a large circle in Figure~\ref{BKetaGG}
can be calculated in the soft $K$ limit within
the HL$\chi$QM, 
as displayed
in Figure~\ref{BKGGst}.

To begin with it, let us recall that the
involved
vector current form factors 
in heavy quark physics are defined by 
\begin{equation}
\begin{split}
 \langle K| \bar{s}\gamma^{\mu} b | B \rangle& = 
f_{+}(q^2) (p_B + p_K)^\mu +  f_{-}(q^2) (p_B - p_K)^\mu \\[1ex]
            & = F_{B}(q^2) M_{B} v^{\mu} + F_{K}(q^2) p_{K}^\mu \;,
\end{split}
\end{equation}
where
\begin{equation}
 F_{B,K} \equiv f_+ \pm f_- \;, \qquad \quad  \text{and} \qquad \quad
p_{B}^\mu = M_{B} v^\mu  \;.
\end{equation}
In the $M_B \to \infty$ limit, $F_{K}$  dominates, as it can be seen from
the scaling properties $F_K \sim \sqrt{M_B}$ and  $F_B \sim
1/\sqrt{M_B}$ \cite{Isgur:1989qw}. In addition, considering  the soft $K$ limit within 
heavy-light chiral perturbation theory (HL$\chi$PT) \cite{HLChPT},
 $F_{K}$ is dominated by the $B_{s}^*$ pole,
and is given by
\begin{equation}
 \big(F_{K}\big)_{\rm soft} =  C_{\gamma} \frac{\sqrt{M_B}}{f_\pi \sqrt{2}}
\frac{g_A \alpha_H}{v\cdot p_K} \;,
\label{BpoleFK}
\end{equation}
where $\alpha_{H}=f_{B}\sqrt{M_B}/(C_\gamma + C_v)$, and $C_\gamma \approx 1$,
$C_v \approx 0$ are the coefficients determined by QCD renormalization of the weak
heavy-light current \cite{Ji:1991prBroadhurst:1991fz}.
The soft $K$ limit is of course unphysical in our case, and to overcome this
we employ a double pole structure
\begin{equation}
 F_{K}(q^2) = 
\frac{F_{K}(0)}{\big(1-\frac{q^2}{M_{1}^2}\big)\big(1-\frac{q^2}{M_{2}^2}\big)} \;.
\label{doublepole}
\end{equation}
Such a structure, proposed in \cite{Becirevic:1999kt},
seems to fit very well the
existing data and the theoretical requirements on the heavy-light
vector form factor, for, say, $M_1 = M_{B_{s}^*}$, 
and some parameter $\gamma$  fitting $M_{2}^{2} = \gamma M_{B_{s}^*}^2$.
To determine $\gamma$ we observe that (\ref{doublepole}) in our limits reads
\begin{equation}
 F_{K}(q^2)_{\rm soft} = \frac{\gamma M_B  F_{K}(0)}{2(\gamma - 1)} 
\frac{1}{v\cdot p_K} \;,
\end{equation}
so that a comparison with (\ref{BpoleFK}) gives
\begin{equation}
 \frac{\gamma F_{K}(0)}{2(\gamma - 1)} =  \frac{C_\gamma}{C_\gamma + C_v}
\frac{f_B}{f_\pi \sqrt{2}} g_A \; ,
\label{lambdafromFK}
\end{equation}
which is general within HL$\chi$PT. 
Knowing the value for $F_{K}(0)$, $\gamma$ can be determined.
We will use the result of the QCD sum rules on the light-cone analysis,
$F_{K}(0)=0.34\pm 0.05$ \cite{Ball:1998tj}, implying $\gamma= 1.27 \pm 0.08$,
in agreement with lattice fits \cite{Becirevic:1999kt}.
 Thus, extrapolating from the
soft $K$ to the general case we obtain the substitution rule
\begin{equation}
 \frac{1}{v\cdot p_K} \quad\longrightarrow \quad
  \frac{2(\gamma -
  1)}{M_B\gamma}\frac{1}{\big(1-\frac{q^2}{M_{B_{s}^*}^2}\big)
\big(1-\frac{q^2}{\gamma M_{B_{s}^*}^2}\big)} 
= \frac{\sqrt{2}f_\pi F_{K}(q^2)}{f_B g_A M_B} \;.
\label{extrapolation}
\end{equation}

Below we will  assume that the form factor for the $B\to K G G^*$ vertex
also has the dipole form (\ref{doublepole}), because of  $B_{s}^*$ pole
dominance in both cases.
 This assumption will
hold within HL$\chi$PT in the region where
it is valid.  Therefore  we  
will adopt the rule (\ref{extrapolation})  also for the $B\to K G G^*$
form factor. For this case, the position of the second pole might be
somewhat different, which means that $\gamma$ and $F_K(0)$ in 
(\ref{lambdafromFK}) should be 
replaced by   $\gamma_G$ and $F_K^G(0)$, respectively.

The $b \rightarrow s G^*$ penguin operator at the quark level is
\begin{equation}
 g_s G_{P} \bar{s}\gamma^\mu L t^a b (DG)^{a}_\mu \;,
\label{Peng}
\end{equation}
where $D$ means a covariant derivative, and
\begin{equation}
 G_P = \frac{G_F}{\sqrt{2}} \frac{1}{4 \pi^2} V^{*}_{ts} V_{tb} (F_{1}^t -
 F_{1}^c) \; .
\label{GP}
\end{equation}
For the quantity $(F_{1}^t - F_{1}^c)$ we take the one loop result \cite{Hou:1987vd},
$0.26 - \big(-(2/3)\ln m_{c}^2 / M_{W}^2\big) \approx -5.2 $.
This result might be slightly changed by perturbative QCD effects
like in \cite{Galic:1980vg} for $s \rightarrow  d G^*$,
but we do not enter such details here.

Note that the contribution of the \emph{dipole} penguin operator
$F_2 m_b \bar{s} \sigma^{\mu\nu} R t^a b G^{a}_{\mu \nu} $ is suppressed
by the small form factor $F_{2}\approx 0.2$, and we neglect it.

The bosonization of the coloured quark current in (\ref{Peng})
 with emission of an additional  soft gluon
is known \cite{Hiorth:2003ci},
\begin{equation}
 \bar{q}_L \gamma^\mu L t^a Q_v  \quad\longrightarrow\quad
G_H g_s G^{a}_{\alpha\beta} \: {\rm Tr} \bigg\{
\xi^\dagger \gamma^\mu L H_v \big[ A_1 \sigma^{\alpha\beta} + 
A_2 \sigma^{\alpha\beta} \gamma\cdot v \big]\bigg\} \;,
\label{BtoG}
\end{equation}
where 
\begin{equation}
H_v  = \frac{1+\gamma\cdot v}{2} \Big( \bar{B}^{*}_{\mu}\gamma^{\mu} - i
 \bar{B}_5 \gamma_5 \Big)
\end{equation}
is the heavy meson ``superfield'' \cite{HLChPT} and  
$\xi=\exp(i\Pi/f_\pi)$, with $\Pi$ being
the standard Goldstone boson
$3\times 3$ matrix field. Furthermore,  $G_H$ is the  meson-quark coupling
given by  $G_{H}^2 = 2 m \rho / f_{\pi}^2 $, where 
$m=0.250\pm 0.025$ GeV is the constituent light quark mass,  $\rho 
\approx 1.1$ is a hadronic parameter \cite{Hiorth:2003ci}  and 
 $g_A = 0.59 \pm 0.08$ is the axial coupling of Goldstone bosons
to heavy mesons.
Moreover, within the HL$\chi$QM we find
\begin{equation}
 A_1=-\frac{1}{8} \bigg(\frac{1}{8 \pi} - i I_2\biggr) \;, \qquad
 A_2=-\frac{1}{8} i I_2 \;,
\end{equation}
and $I_2$ is a logarithmically divergent loop integral which
 is expressed in terms  of $f_\pi$ and
the gluon condensate  \cite{ERT,PideR,BEF,Hiorth:2002pp}, as follows
\begin{equation}
 f_{\pi}^2 = - 4 i m^2 N_c I_2 +
 \frac{1}{24 m^2} \langle \frac{\alpha_s}{\pi}G^2 \rangle \;.
\end{equation}
 It should be noted that the  structure in (\ref{BtoG}) should be rather general,
  while $G_H$ and the explicit expressions for $A_{1,2}$ are model dependent.

Taking the vector ($B^*$) part of the $H_v$ and connecting with the $B_{s}^*$ propagator
in Figure~\ref{BKGGst}, we obtain in the soft $K$ limit the amplitude
\begin{multline}
 M(B\to K G G^*)_{\rm soft} = i \frac{g_A \sqrt{M_B}}{f_\pi
 \sqrt{2}}\:
 G_H\: G_P\: g_s\: 
G_{\alpha\beta}^a \: (DG)^{a}_\mu \: \frac{(p_K)_\nu}{(v \cdot p_K)} \\[1ex]
  \quad\times \Big\{
-(A_1+A_2)\epsilon^{\mu\nu\alpha\beta} + 
2 A_2 v^\mu \epsilon^{\lambda\nu\alpha\beta}v_\lambda 
\Big\} \;.
\label{softamp}
\end{multline}
In order to obtain the general amplitude for $B\to K G G^*$
from this equation,
we perform the substitution (\ref{extrapolation})
for  $(v \cdot p_K)$ in the denominator above with $F_K(q^2)$ replaced
by  $F_K^G(q^2)$.

Concerning
the $\eta' G^* G'$ interaction, 
it has the general form already used in ref. \cite{Eeg:2001tg},
\begin{equation}
 V(G^* \to G' \eta') = - \frac{1}{2} F_{\eta' g g}(q^2) \delta^{a' c}
\epsilon^{\rho\lambda\kappa\sigma}\epsilon_{\rho}^{*a'} G^{c}_{\lambda\kappa}
q_\sigma \;,
\end{equation}
and several groups \cite{Muta:1999tcKroll:2002nt,Ali:2000ci}
calculated the form-factor $F_{\eta' g g}(q^2)$ in the perturbative QCD
approach. Since this approach becomes unreliable for gluon momenta
of $\sim 1$ GeV, we adopt
a formula from \cite{Ali:2000ci} which interpolates
between the perturbative QCD region and the anomaly value for zero momentum. 
This formula gives $F_{\eta' g g}(m_{\eta'}^2)=1.55 \pm 0.40$ GeV$^{-1}$, where an
error of 25 \% has been
allowed. Accordingly, together with the value of the gluon condensate,
this is the major source of uncertainty in our result below.
Taking now the vacuum expectation value of the two soft gluons
\begin{equation}
 g_{s}^2 G^{a}_{\alpha\beta}G^{c}_{\lambda\kappa} \quad\longrightarrow\quad
\frac{4 \pi^2 \delta^{ac}}{12(N_c^2-1)}\Big(g_{\alpha\lambda} g_{\beta\kappa}-
g_{\alpha\kappa} g_{\beta\lambda}\Big)\langle \frac{\alpha_{s}}{\pi} G^2 \rangle \;,
\end{equation}
we obtain the final amplitude
\begin{equation}
 M(B\to K \eta')_{\langle G^2 \rangle} = 
 \frac{ \pi^2  G_P  G_H  F_{K}^G(q^2) \, F_{\eta' g g}(q^2)}{3 f_B \sqrt{M_B}} 
\langle \frac{\alpha_s}{\pi}G^2 \rangle
 M_{B}^2 \Big\{ A_2- 3(A_1+A_2)\Big\} \;,
\label{physamp}
\end{equation}
where $q^2 = m_{\eta'}^2$.

Numerically, with 
$\langle \frac{\alpha_s}{\pi}G^2 \rangle = (0.32\pm 0.02\; {\rm GeV})^4$, 
$f_{B}= 0.18 \pm 0.03$ GeV, and $F_K^G(q^2)=F_K(q^2)$ given by the numbers
below Eq. (\ref{lambdafromFK}), we get
\begin{equation}
 |M(B\to K \eta')_{\langle G^2 \rangle}| = (8 \pm 3) \times 10^{-9} \;
  \text{GeV} \; .
\label{eight}
\end{equation}
It should be noted, that even if the uncertainty of $\gamma$ is
increased to 50 \% when replaced by $\gamma_G$, it has no significant impact on
the form factor $F_K^G(q^2)$ at the physical point $q^2 =
m_{\eta'}^2$, and thereby not on the final
 result (\ref{eight}). This means that our assumption for the form
factor $F_K^G(q^2)$ should be rather sound.

Our result (\ref{eight})
 should be compared to the experimental amplitude $|M(B\to K \eta')_{\rm exp}| =
(88 \pm 2) \times 10^{-9}$ GeV. 
Thus, according to our analysis,
 only of order 10$\%$
 of the $B\to K \eta'$ rate enhancement can be ascribed 
to this gluonic creation of $\eta'$. Some additional mechanisms are
necessary, 
such as constructive
interference of amplitudes for creating $\eta'$ in $d\bar{d}$ and $s\bar{s}$ state
\cite{Lipkin:1990us,Beneke:2002jn}.
However, the gluonic mechanism studied here, with its distinctive flavour-singlet nature,
seems to go in the direction of the result of the SU(3) symmetry
analysis in  \cite{Chiang:2003rb}
that shows a substantial singlet penguin contribution to $B\to K \eta'$
amplitude.

\section{Discussion}

 Let us comment here on how our result fits into the existing accounts of
joining the two-gluon  configurations arising from the 
$b\to s G$ and $b\to s  GG$ transitions with those inherent to the $\eta'$ particle,
in particular on those relying on properties of the $\eta'$ particle that
are related to the QCD anomaly.
The attempts to explain some puzzling hadronic weak decays by QCD anomalies
are well known: the $\Delta I = 1/2$
enhancement in $K\to \pi\pi$ by the trace anomaly \cite{PeP94GeW00} and the
enhancement of $B\to \eta' X_s$ decay rates by the axial
anomaly \cite{bketatheor,Fr97,GerTr}.

We entered such study by employing the fact that anomaly permeates all
distance scales, that enables one to study the role of two-gluon anomalous
configurations
from an extreme SD to a truly LD regime. Our recent study \cite{EeKP03}
shows that there is merely a remnant, the anomaly
tail, in the extreme SD case. 
It should be noted that this contribution is obtained as a two quark
operator for $b \rightarrow s \eta'$ and is very different from the LD
contribution presented in detail in the previous section.

In addition, we have also identified some
other contributions which we have found to be negligible.
For example, Fritzsch \cite{Fr97} has suggested that an effective
interaction of the form 
\begin{equation}
H_{\text{eff}} = a \alpha_s G_F \bar{s}_L b_R G_{\mu\nu}
\tilde{G}^{\mu\nu} 
\end{equation}
might contribute significantly. 
We will describe elsewhere \cite{EKP-bigger}
different perturbative and nonperturbative contributions to
such an effective  interaction stemming from anomalous two-gluon configurations.
 Note that already a
rotation of an appropriate term from Simma and Wyler's paper \cite{SiW90}
to Fritzsch's form enables one to read off a
perturbative contribution to coefficient  $a$ above, 
 $a^{\text{pert}}_{\text{SW}} \simeq 4 
 \times 10^{-4} \;\text{GeV}^{-1}$, which gives an amplitude 3--4 times
smaller than (\ref{eight}).

After observing a systematic suppression of the anomalous two-gluon
contributions through all the distance scales \cite{EKP-bigger}, 
we are focusing in the present paper to LD
nonperturbative gluon configurations that may 
be phenomenologically more relevant.
Thereby we are considering the low energy contribution where a
gluon condensate accounts for the emission of soft gluons. A priori, our
calculation would only be valid in the unphysical case where the outgoing
kaon is soft. However,
one can extrapolate to the physical point by
introducing a 
dipole form factor for the $B \rightarrow K GG^*$ transition (as for
the standard  $B \rightarrow K$ transition current).
As a result we find a more significant 
contribution from this mechanism, which can account for $\sim 10 \%$
of the measured amplitude.

This result has to be compared with findings of
\cite{Beneke:2002jn} that flavour singlet contributions to $B\to K \eta'$
may be marginal.
However, due to quite large uncertainties in both amplitudes
these two results are actually not inconsistent.
Note that the major portion of the singlet penguin amplitude in \cite{Beneke:2002jn}
comes from the operators corresponding to singlet quark configurations
forming $\eta'$ particle.
Recently another analysis within SCET appeared \cite{WilZup}. This
analysis concludes that the ``singlet penguin'' contribution is
essential to understand the process $B \rightarrow K \eta'$.
Unfortunately, a direct comparison between our treatment of one soft ($q^2 \sim 0$)
and one semi-hard ($q^2 \sim m_{\eta'}^2$) gluon and the one by SCET is difficult to perform.
Anyway,
our contribution corresponding to the gluonic
configurations 
forming the  gluon condensate 
is a novel one, and may significantly increase the
role of the singlet penguin mechanism in the direction of the result based 
on the SU(3) symmetry analysis \cite{Chiang:2003rb}.

A number of authors already used the surprising 
$B\to K \eta'$ enhancement to infer on the contributions beyond the
Standard model.
 However, at this stage we need first to consider possible contributions from the
specific mechanisms within the Standard model, like the one presented here. 
This mechanism seems to provide an additional contribution to the
singlet penguin topology, the understanding
of which may 
be of importance for explaining the data on CP asymmetries in penguin
dominated modes \cite{HFAG}.

\subsubsection*{Acknowledgment}
K.~K. and I.~P. gratefully acknowledge the support of the Norwegian Research Council 
and the hospitality of the Department of Physics in Oslo, as well as support of
Croatian Ministry  of Science, Education and Sport under the contract No. 0119261.
J.~O.~E.  is supported in part by the Norwegian
research council and  by the European Union RTN
network, Contract No. HPRN-CT-2002-00311 (EURIDICE). 
K.~K. and J.~O.~E. thank Jure Zupan for fruitful discussions about
the treatment of the considered process within SCET.

\end{document}